\documentclass[journal]{IEEEtran}
\usepackage{color}
\usepackage{cite}
\usepackage{amsmath,amssymb,amsfonts}
\usepackage{graphicx}
\usepackage{textcomp}
\usepackage{multirow}
\usepackage{amsthm}
\usepackage{mathrsfs}
\usepackage{xcolor}
\usepackage{manyfoot}
\usepackage{booktabs}
\usepackage{algorithm}
\usepackage{algorithmicx}
\usepackage{listings}
\usepackage{hyperref}
\usepackage{soul}

\graphicspath{{./final_figures/}}

\def\BibTeX{{\rm B\kern-.05em{\sc i\kern-.025em b}\kern-.08em
    T\kern-.1667em\lower.7ex\hbox{E}\kern-.125emX}}

\begin{document}
\title{
Highly uniform first-electron position in qubit arrays fabricated on dedicated QSOI\textsuperscript{\textsuperscript{\textregistered}} 300mm commercial platform
}
\author{Johan Pelloux-Prayer,
    Elise Prin, Giselle A. Elbaz, Pierre-Louis Julliard, Amaryllis Comiti, Clément Nguyen, Sylvain Martin, Patrick Torresani, Renan Lethiecq, Carlos Augusto Suarez Segovia, Franck Arnaud, Etienne Nowak, Tristan Meunier, Bruna Cardoso Paz
\thanks{This work was financially supported by the BPI iDémo project Q100T (grant number DOS0254912 and DOS0254913), the Proqcima initiative, both under the France 2030 program, EIC Transition project MCSquare (grant number 101136414)) and JU CHIPS project ARCTIC (Grant number 101139908).}
\thanks{Johan Pelloux-Prayer,
    Elise Prin, Giselle A. Elbaz, Pierre-Louis Julliard, Amaryllis Comiti, Clément Nguyen, Sylvain Martin, Patrick Torresani, Renan Lethiecq, Etienne Nowak, Tristan Meunier and Bruna Cardoso Paz are with Quobly, Grenoble, France. (e-mail: bruna.cardoso-paz@quobly.io).}
\thanks{Carlos Suarez Segovia and Franck Arnaud are with STMicroelectronics, Crolles, France.}}

\maketitle

\begin{abstract}

We report progress toward the development of a quantum silicon-on-insulator (QSOI\textsuperscript{\textregistered}) technology compatible with 300mm CMOS fabrication and adapted from the 28nm Fully-Depleted SOI (28nm FD-SOI) platform for scalable quantum computing.
We compare quantum devices fabricated with standard 28nm FD-SOI and QSOI\textsuperscript{\textregistered} technologies, and show striking improvements of room temperature electrostatic properties of the individual device.
Moreover, the QSOI\textsuperscript{\textregistered} technology has significantly reduced the device variability and the dispersion of device properties at the wafer level.
Transistor metrics are reproduced by TCAD simulations showing that the electrostatics of the devices behave as expected for QSOI\textsuperscript{\textregistered} technology.
Wafer-scale measurements at sub-2K show reproducible quantum dots down to the few-electron regime with 69\% yield for successful charge detection of the first electron and a dispersion of the first electron position of $\mathrm{\pm 35 mV}$ over 377 quantum dots.
These results establish QSOI\textsuperscript{\textregistered} as a promising platform for CMOS-compatible quantum device co-integration.
\end{abstract}

\section{Introduction}
\label{sec:introduction}

The development of silicon-based quantum technologies has gained significant momentum due to their compatibility with
CMOS fabrication processes and therefore their potential for large-scale integration~\cite{gonzalez_zalba_scaling_2021, maurand_cmos_2016, zwanenburg_silicon_2013}.
Among the various quantum computing platforms, silicon spin qubits have emerged as a promising candidate, offering long
coherence times and the possibility of monolithic integration with control/readout electronics~\cite{meunier_silicon_2025, dumoulin_stuyck_cmos_2026, hamonic_foundry-fabricated_2025}.
However, the path toward scalable quantum processors requires not only high-fidelity qubit operation but also
reproducible device performance across wafers~\cite{neyens_probing_2024}.

Recent efforts have demonstrated single- and two-qubit operations in different Si platforms, including 28nm FD-SOI, SiMOS
and SiGe heterostructures~\cite{bartee_spin_2024, cardoso_paz_fdsoi_2024, zajac_resonantly_2018, scappucci_germanium_2020, veldhorst_two_qubit_2015}.
However, these devices are often still fabricated in research-scale environments, where process control and
wafer-level variability analysis remain limited.
For industrial quantum technologies, commercial CMOS platforms offer the low variability, high yield, and process uniformity needed to achieve a reliable product at the $\mathrm{300mm}$ scale
~\cite{neyens_probing_2024}.
Moreover, as qubit arrays continue to scale up, wafer-scale electrical characterization at both room and cryogenic
temperatures becomes increasingly important for quality control, process optimization,
and pre-dicing device performance characterization.

To address these challenges, we are collaborating with STMicroelectronics to develop the quantum version of their 28nm FD-SOI
proprietary process-of-record - the QSOI\textsuperscript{\textregistered} process flow - by carefully isolating and fine-tuning the steps which
negatively impact qubit performance.
In this paper, we present the latest results of devices fabricated using the optimized QSOI\textsuperscript{\textregistered}
process at the wafer scale, combining automated testing at 300K (section ~\ref{subsec:MOS}) and sub-2K
(section~\ref{subsec:wafer_level_cryogenic_characterization}).
We probe a bilinear array that is our 2 qubit-gate unit cell at sub-2K, where each gate is tuned to operate
both in the many-electron regime (as a charge sensor) and in the few-electron regime (as a quantum dot, QD).
Our results demonstrate wafer-scale yield and a low variability in the position of the first electron
in QSOI\textsuperscript{\textregistered} devices.
This marks a critical milestone in the validation of our CMOS-compatible quantum technology platform.

\section{Device and Integration Details}
\label{sec:device_and_integration_details}

The devices investigated in this work are linear and bilinear arrays
fabricated using our QSOI\textsuperscript{\textregistered} CMOS-compatible quantum platform.
In this platform, the QDs are electrostatically defined by plunger gates (G) deposited on top of SOI silicon nanowires defined by shallow trench isolation.

\begin{figure}[!t]
\centerline{\includegraphics[width=\columnwidth]{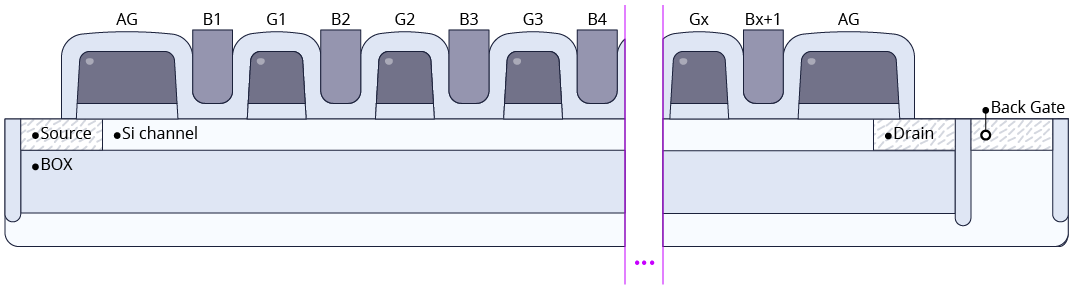}}
\caption{Schematic of an array of quantum dots using QSOI\textsuperscript{\textregistered}. Access gates (AG) are used to form electrostatic reservoirs. Plungers gates (G1, G2,..., Gx) are used to control the chemical potential of the quantum dots. Barriers (B1, B2, ..., Bx+1) are repurposed vias used to control the tunnel coupling between two neighboring quantum dots. This is an artist impression and does not reflect the real dimensions of our devices.}
\label{fig:QSOI_schematic}
\end{figure}

As previously described in~\cite{elbaz_integration_2025}, we integrate additional tungsten vias in between the plunger gates as a second level of gates (B) - namely barrier gates - and control the coupling between neighboring QDs.
Finally, as the arrays are built on SOI, we have an additional knob, the back gate (BG), which can electrostatically affect the position and quality of the created QDs.
Through carefully designed DOEs (Design of experiments) and process development, including adjustments to the gate stack, intergate stack and critical dimensions, we have significantly improved the transfer characteristics of our qubit arrays and greatly reduced the device-to-device variability.

\section{Methodology and setup}\label{sec:methodology-and-setup}
The electrical characterization flow followed a straightforward screening logic, moving the samples from the highest to
lowest throughput measurement setups.
Each lot went through fully automated electrical probing at 300K\@.
Transistor-level metrics such as those presented in section ~\ref{subsec:MOS} were collected.
In addition to the qubit devices, test structures like transistors and capacitors were measured for process monitoring.
The purpose of this room-temperature testing is to provide a first-level statistical screening that enables the
ranking of different process splits and the down-selection of the most appropriate wafers for further cryogenic studies.

Selected wafers of the best lot/split were cooled down using a semi-automatic wafer-level cryogenic probestation.
Although the chuck is operated at about 900mK, the wafer temperature is estimated to be closer to 2K by the tool supplier\@.
In addition to the Id-Vg curves for the MOSFET regime, DC charge properties for the
quantum dots were also measured and are presented in sec.~\ref{subsec:wafer_level_cryogenic_characterization}.

We probed 7098 and 2615 devices at room and cryogenic temperatures, respectively.
The drastic decrease in throughput at cryogenic temperatures is both due to the thermalization time required to reach base
temperature ($\mathrm{150min / 300mm}$ wafer) and the amount of data collected at 2K (2D maps) versus 300K (1D traces only).
Using custom algorithms, we extracted key MOS and QD metrics to differentiate the different splits.
To complement experimental characterization, we used TCAD simulations to confirm that device electrostatics are as expected and well understood.

\section{Results and Discussion}\label{sec:results_and_discussion}
\begin{figure*}[!ht]
 \centering
\includegraphics[width=0.7\textwidth]{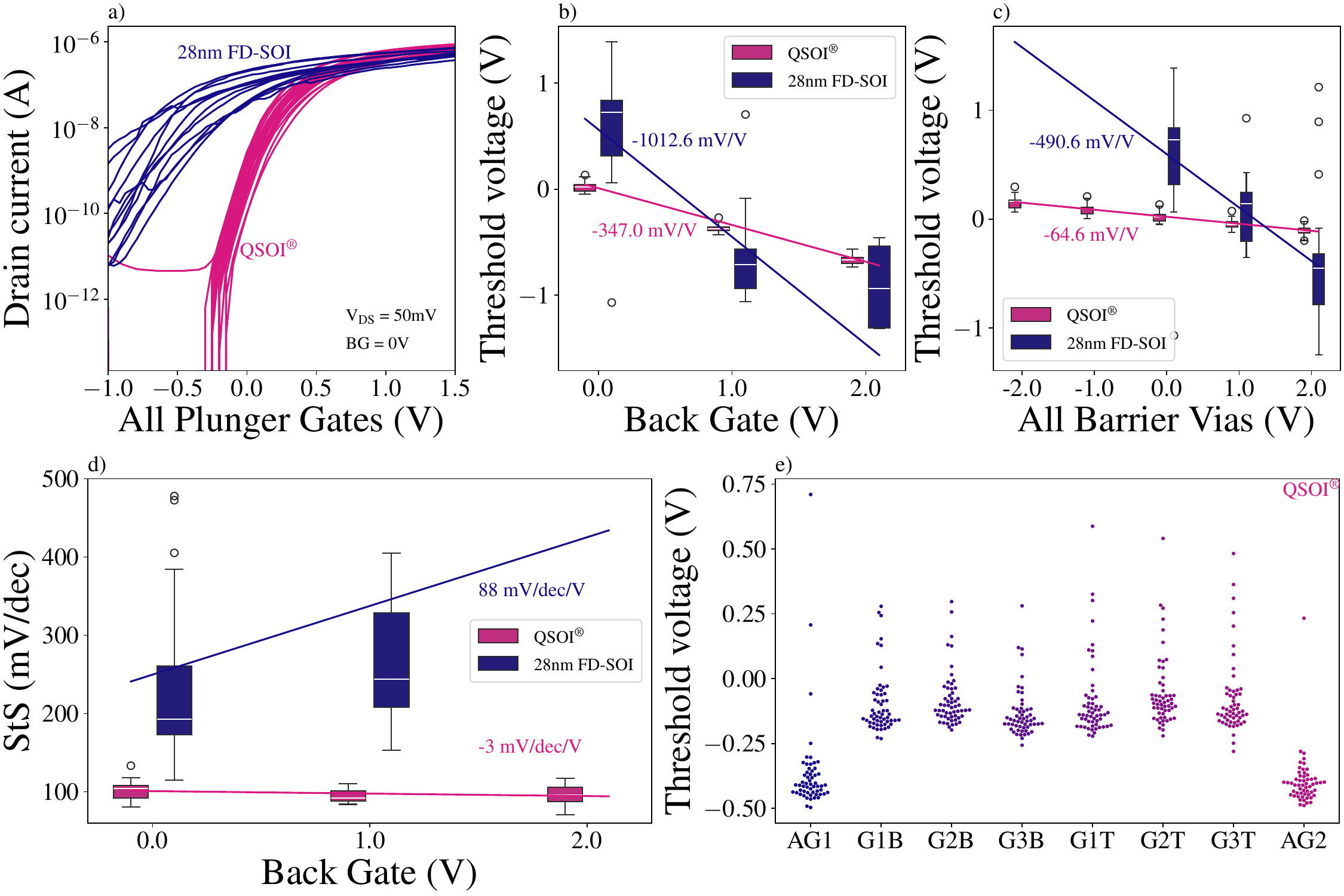}
\caption{
Data are colored in blue for 28nm FD-SOI and in pink for QSOI\textsuperscript{\textregistered} in a, b, c, and d.
a) Wafer-level Id-Vg across the center, mid and edge radii (28 dies).
b) - c) Boxplot of the threshold voltage for 28nm FD-SOI and QSOI\textsuperscript{\textregistered} against the applied back-gate voltage in (b) and the barrier-vias voltage in (c).
 Box plots show the interquartile range and the median.
 Whiskers represent the maximum and minimum values without outliers.
 Outliers are defined as data points that are more than 1.5 interquartile range away from the median.
d) Boxplot of the subthreshold swing (StS) extracted from Id-Vg curve for 3 back gate biases: 0V, 1V, and 2V\@. For a back gate of 2V the 28nm FD-SOI data is shifted to the negative, outside of our measuring range, which prevents the extraction of the subthreshold swing.
e) Swarm plot showing the QSOI\textsuperscript{\textregistered} threshold voltage distribution for each individual plunger gate in a 2x3 qubit array.
 Data shown is for QSOI\textsuperscript{\textregistered} measurements only.
}
\label{fig:room_temperature}
\end{figure*}

The following section presents MOS and QD electrical characterization results obtained at wafer-level scale.
MOSFET data was collected for bilinear arrays with devices operating in the MOS regime at $\mathrm{V_{DS}}$ = 50mV\@.
QD data was collected for bilinear arrays with devices operating in the quantum regime at $\mathrm{V_{DS}}$ = 1mV for the best-performing process split.
Sections~\ref{subsec:MOS} and~\ref{subsec:wafer_level_cryogenic_characterization} show wafer-level measurements that
illustrate, respectively, both the uniformity and the yield of the few-electron regime.

\subsection{Wafer-level MOS characterization}
\label{subsec:MOS}

To obtain wafer-level statistics of the electrical properties for each of our lots, we measured our bilinear arrays in 28 dies distributed across the center, mid and edge radius of our wafers.
Fig.~\ref{fig:room_temperature}a demonstrates the stark improvement brought by the QSOI\textsuperscript{\textregistered} technology compared to 28nm FD-SOI on a 2x3 bilinear array measured as a traditional MOSFET (sweeping all plunger gates together) at room temperature.
As can be seen, the bilinear arrays built with 28nm FD-SOI technology show a very negative threshold voltage ($\mathrm{V_{TH}}$), with a large dispersion of $\mathrm{\sigma_{V_{TH}} = \mathrm{385 mV}}$.
In comparison, the same bilinear array built with QSOI\textsuperscript{\textregistered} technology shows tightly grouped curves for the 28 dies measured with $\mathrm{\sigma_{V_{TH}} = 41.1 mV}$.

Whereas the 28nm FD-SOI technology performance is on par with state-of-the-art for standard transistors~\cite{mazurier_variability_2014}, the degradation we observe here is due to the specific design of our qubit devices, which deviates from standard transistors and needs to be optimized to recover the expected performance.
Fig.~\ref{fig:room_temperature}b and~\ref{fig:room_temperature}c show, respectively, how the back gate and the barrier vias are able to tune the conduction inside the structure.
Fig.~\ref{fig:room_temperature}b and~\ref{fig:room_temperature}c show, respectively, how the back gate and the barrier vias are able to tune the conduction inside the structure.
We extract a tuning factor from these measurements which quantifies how a potential applied on the back gate or the vias is able to change the position of the threshold voltage.
We report an average tuning factor value of $\mathrm{-347 mV/V}$ for the back gate and of $\mathrm{-64.6 mV/V}$ for the barrier vias for QSOI\textsuperscript{\textregistered} technology.
The barrier vias have a lower tuning factor comparing to the back gate because they have a lower capacitive coupling with the channel due to screening from the plunger gates.
In comparison, the standard 28nm FD-SOI technology tuning factor is $\mathrm{-1013 mV/V}$ for the back gate and $\mathrm{-491 mV/V}$ for the barrier vias for our devices.
These stronger values can be explained by the standard 28nm FD-SOI plunger gates having poor electrostatic control of the Si channel in our design, allowing the back gate and barrier vias to have a larger impact.
The delicate interplay between the various electrostatic components is important, allowing us to carefully optimize the quantum dot behavior at cryogenic temperatures.
We have found that each of these functionalities at 300K is a strong indicator of their behavior and utility at 2K and below.
Fig.~\ref{fig:room_temperature}d shows the improvement of the subthreshold swing value for QSOI\textsuperscript{\textregistered} vs.\ 28nm FD-SOI\@.
We observe strong improvement in both average value and dispersion of the subthreshold swing, reducing them from $\mathrm{193 \pm 155 mV/dec}$ for 28nm FD-SOI to $\mathrm{104 \pm 10 mV/dec}$ for QSOI\textsuperscript{\textregistered}\@.
Fig.~\ref{fig:room_temperature}e presents the threshold voltage distribution of each plunger gate of our QSOI\textsuperscript{\textregistered} 2x3 bilinear array.
We can see the internal plunger gates (denoted by GxT or GxB for top side or bottom side, respectively) display good matching threshold voltages, where the largest difference in median $\mathrm{V_{TH}}$ observed is about $\mathrm{68.5 mV}$.
This can be compared to previously reported values in~\cite{neyens_probing_2024}, where the standard deviation of the matched-pair $\mathrm{\Delta V_{TH} / \sqrt{2}}$ distribution is 59mV for~\cite{neyens_probing_2024} and 101mV over two wafers in this work.
We believe it is possible to improve the variability of the devices by further optimizing some specific building blocks of the process, like the recently repurposed barrier vias introduced to control the tunnel coupling between dots ~\cite{elbaz_integration_2025}.
It is worth noting that the access gates have a noticeably lower $\mathrm{V_{TH}}$ compared to the plunger gates.
It is common for the outermost gates to exhibit lower threshold voltage in such arrays~\cite{neyens_probing_2024}, and this can be explained by the proximity of the gate to the doped source and drain.
This doping can diffuse sufficiently close to the outermost gate, thereby affecting its threshold voltage significantly.

\begin{figure}[!h]
\centerline{\includegraphics[width=\columnwidth]{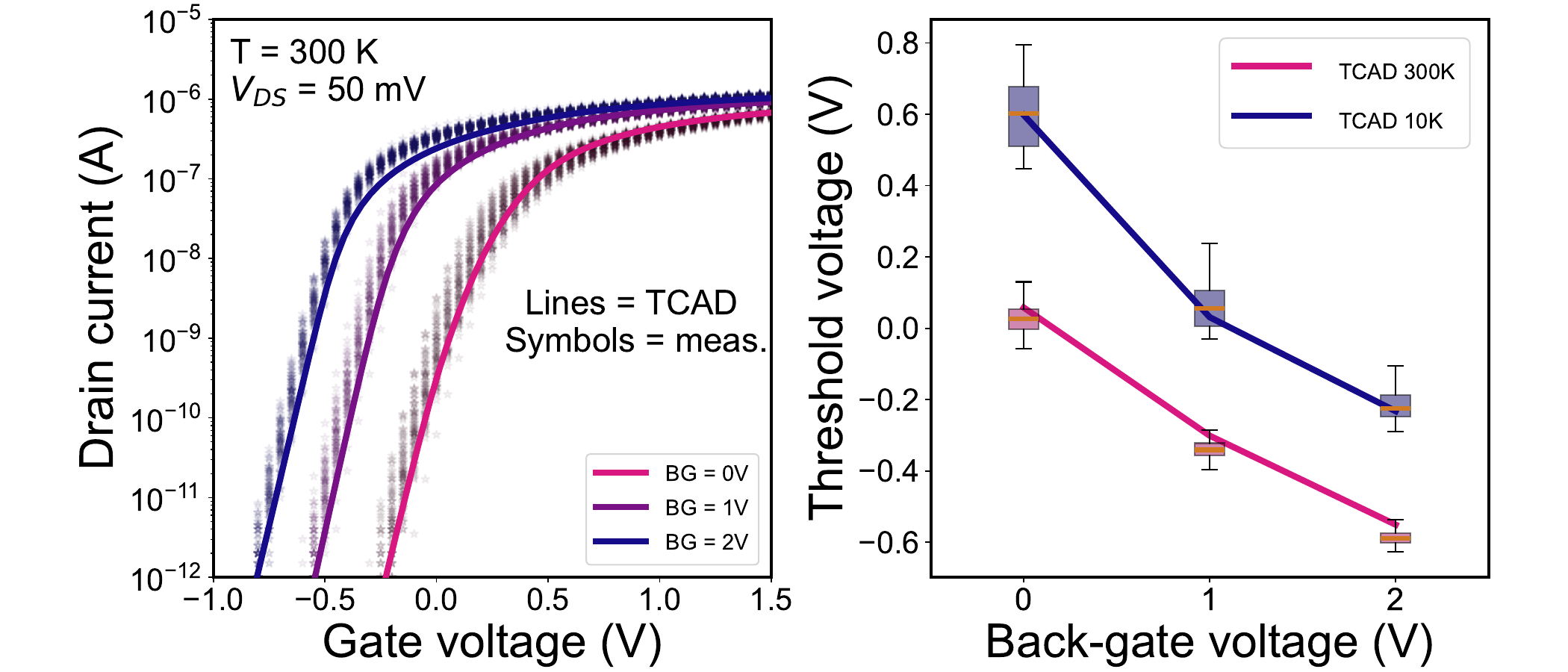}}
\caption{Simulation and experimental data obtained for a linear array with 3 plunger gates in series. Left: transfer characteristics at room temperature for different back-gate bias $V_{BG}$ = 0, 1, and 2V, and $V_{DS}$ = 50mV. Right: Extracted threshold voltage as a function of the back-gate voltage at room and low temperatures.}
\label{fig:simulation}
\end{figure}

Figure ~\ref{fig:simulation} shows good agreement between simulation and experimental data for Id-Vg curves at 300K, and for the $\mathrm{V_{TH}}$ at both room and cryogenic temperatures.
The simulations are calibrated with the result of the electrical characterization of standard transistors with gate dimensions in the micrometer range.
They accurately capture the differences in threshold voltage observed for the various back-gate voltages at room temperature.
The nonlinearity of the body-bias effect at low temperatures is also well reproduced by the simulations.

All these results show that the QSOI\textsuperscript{\textregistered} product line has enabled the fabrication of bilinear arrays that function properly as MOSFETs at room temperature and passes the first screening step toward a functional quantum device at mK temperature.

\subsection{Wafer-level QD characterization}
\label{subsec:wafer_level_cryogenic_characterization}

\begin{figure}[!h]
\centerline{\includegraphics[width=\columnwidth]{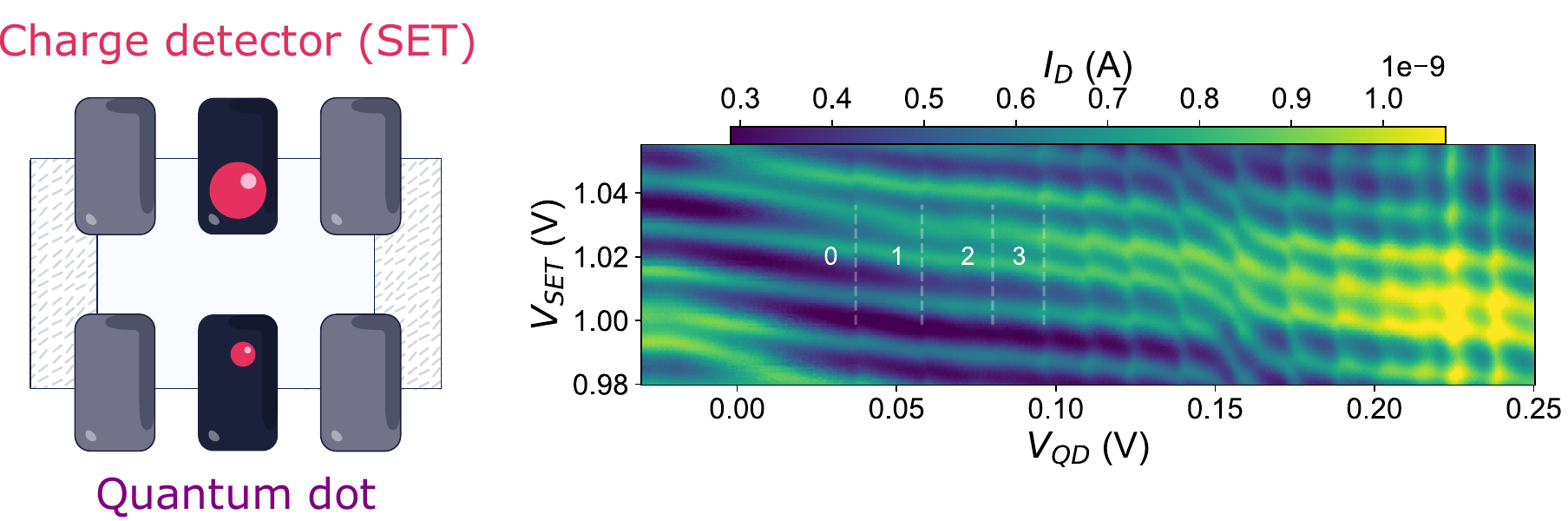}}
\caption{
Left: Schematics of a charge detection configuration on a bilinear device, where a single-electron transistor is used to sense the electrons tunneling inside the quantum dot. Right: Stability diagram showing the few-electron regime for the probed quantum dot formed underneath the plunger gate. The numbers inside the figure indicate the charge occupation of the probed QD\@. The interruption of the regular pattern of vertical jumps indicates that the dot has been emptied.
 Access gates are not shown for simplicity.
}
\label{fig:stability_diagram_few_electron}
\end{figure}

To measure the quantum dot uniformity, each of the 6 plunger gates of the bilinear device was tested both as a charge detector and as a quantum dot (see Fig.~\ref{fig:stability_diagram_few_electron}).
To do this, we operate each plunger gate on one side of the channel in the many-electron (SET or single-electron transistor) regime while operating the plunger gate directly facing it, on the other side of the channel, in the few-electron (QD) regime.
An example of this is shown in Fig.~\ref{fig:stability_diagram_few_electron} left for the center two plunger gates, in one of the two possible SET-QD configurations for this pair.
When in the many-electron regime, the SET can be capacitively coupled to a QD (i.e.\ qubit) and used to read out information about the QD; the drain current measured across the SET yields Coulomb peaks that are shifted each time an electron enters the QD\@.

\begin{figure}[!h]
\centerline{\includegraphics[width=\columnwidth]{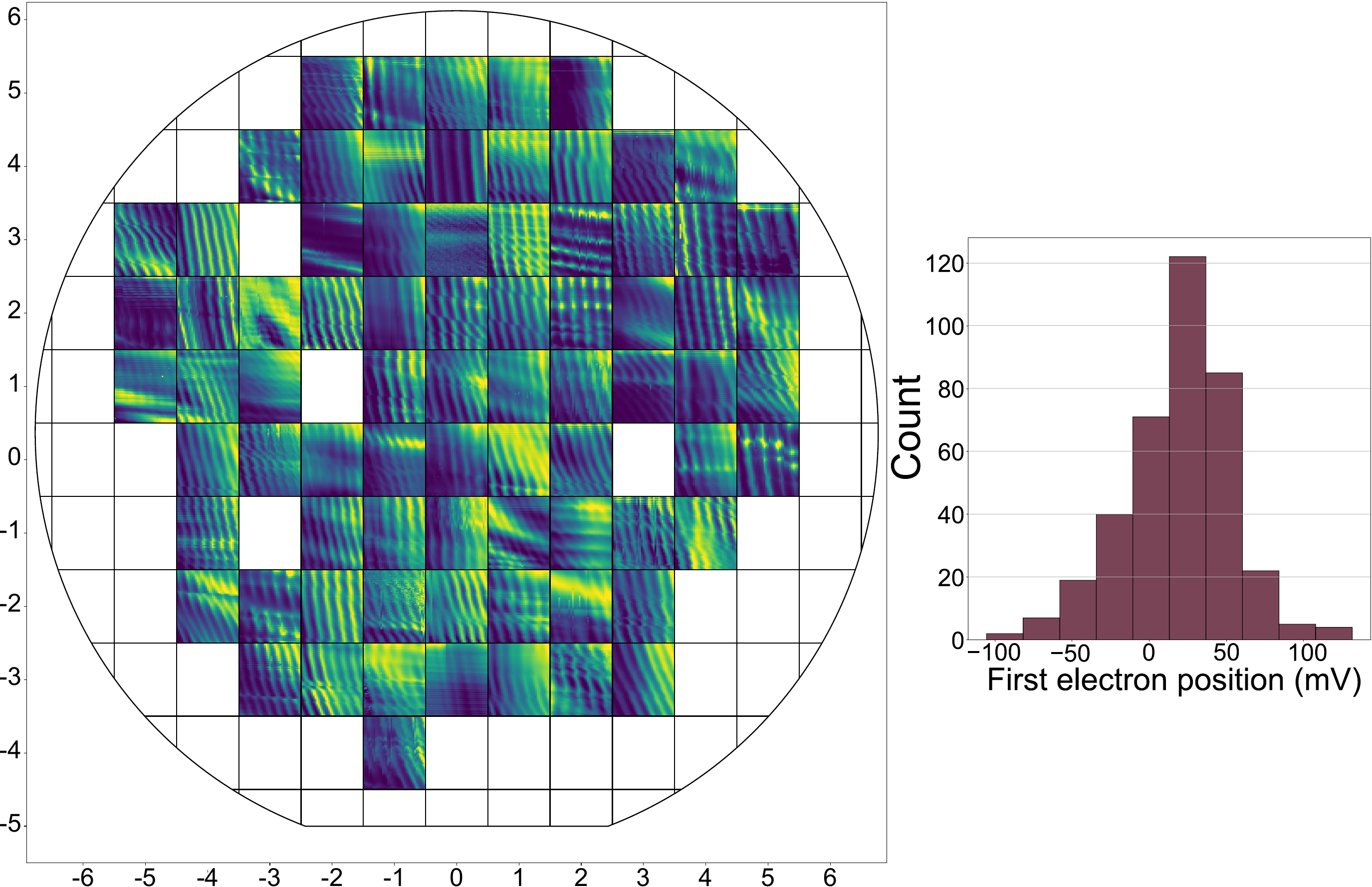}}
\caption{Left: Wafermap of charge detection at the wafer scale for a pair of facing plunger gates. Only 4 dies were not measured since our protocol to find the first Coulomb peak rejected these dies. Each die shows the measurement between plunger gates G3T and G3B\@.
Right: Histogram of the first electron position across the wafer measured for all the possible detectors in our device.}
\label{fig:wafermap_charge_detection}
\end{figure}

The voltage range where an SET is capable of clearly detecting charges entering an opposing QD varies from device to device and even from gate to gate.
To probe a whole wafer, we need to tune the SET voltage range around the first detectable Coulomb peak ($\mathrm{V_{first-peak}}$) for each plunger gate such that $\mathrm{V_{first-peak} + 75mV < V_{SET} < V_{first-peak} + 150mV}$\@.
For the QD, we used a 500mV gate voltage range centered around 0V\@ ($\mathrm{-250mV < V_{QD} < 250mV}$\@).
Figure~\ref{fig:wafermap_charge_detection} left shows an example of wafermap of stability diagrams collected for 91 dies using the above mentioned voltage ranges.
As these measurements were collected for each of the 6 plunger gates, a total of 546 stability diagrams were taken per wafer.
From these, we can define an overall gate yield for this device where the few-electron regime is achieved and find that 69\% or 377/546 of the maps show a detectable first-electron.
This result shows that the process modifications we are making for QSOI\textsuperscript{\textregistered} are moving in the right direction and we can achieve a high yield for our qubit arrays.
There is, of course, still room for improvement both on the device and measurement sides.
For example, to improve the capacitive coupling between SET and QD, and therefore to increase the charge detection signal, a design modification to reduce the distance between the SET detector and QD sides of the nanowires could be considered.
Moreover, the measurement routine does not optimize the detector sensitivity which could further improve the yield.
Additional optimization of the fabrication process could also increase the yield of fully functional devices by reducing variability.

Figure~\ref{fig:wafermap_charge_detection} right shows the gate voltage position for the first electron ($V_{first-electron}$) obtained for the 69\% of devices where this regime is achieved.
The histogram combines the results for each plunger gate of all the bilinear arrays measured.
The first electron position ($\mathrm{V_{first-electron}}$) over all gates and maps is 21mV ± 35mV\@.
To the best of our knowledge this the lowest variability ever observed for the first electron position at wafer-scale ~\cite{neyens_probing_2024}.
The small standard deviation indicates that the $\mathrm{V_{QD}}$ range (here 500mV) can be significantly optimised (reduced) in future measurements, which would decrease the number of data points for the stability diagrams and improve the measurement throughput accordingly.
More importantly, this indicates very low variability and reproducible electrostatics across the wafer, which will facilitate the development of gate voltage calibration routines~\cite{zwolak_data_2024} as well as the fabrication of large qubit arrays with reproducible qubit characteristics.

\begin{figure}[!h]
\centerline{\includegraphics[width=\columnwidth]{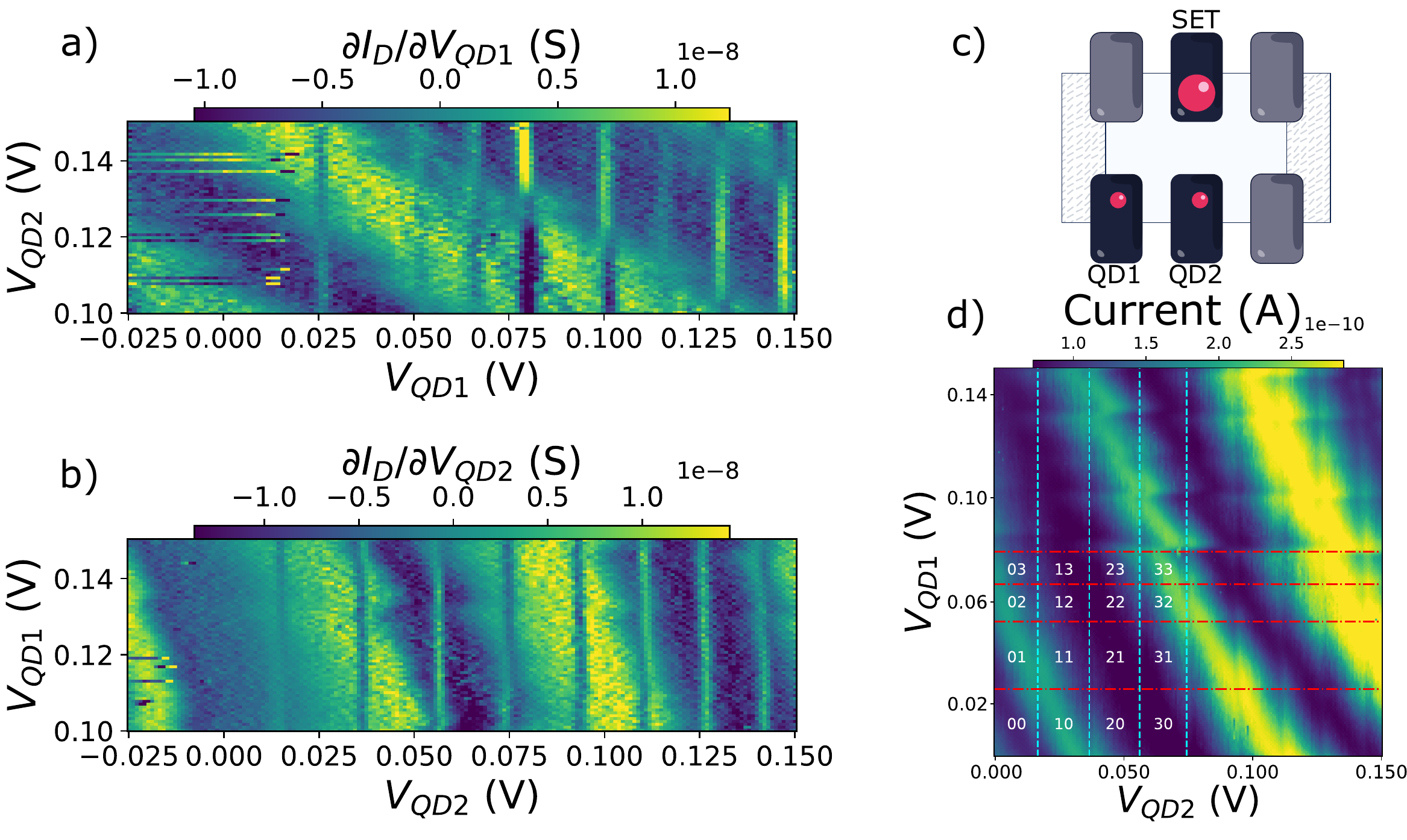}}
\caption{
a) - b) Drain current derivative in the QD1 (QD2) direction. The faster voltage sweep is along the QD1 (QD2) gate to emphasize charge detection along this gate.
c) schematic of the studied bilinear array with the SET and 2 quantum dots.
d) charge detection showing the position of the first electrons for each of the quantum dots located underneath plunger gates G1B and G2B. The numbers inside the figure indicate the charge occupation of the probed double QD\@.
}
\label{fig:doubleQD}
\end{figure}

Once the position for the first electron was obtained for each plunger gate, we moved one step further and collected 2D maps for the charge detection of a double QD (DQD).
To do so, we tuned the SET position to the maximum conductance (derivative of the current with respect to the SET gate voltage), and swept the DQD gates around 0V\@.
An example of one such map is presented in Fig.~\ref{fig:doubleQD}, where the dots are formed underneath two consecutive plunger gates (G1B and G2B), and the SET is located under G2T, the plunger gate facing QD2\@.
This was acquired with 0V applied at the barrier gates and so the 2 QDs are weakly coupled to each other.
Dedicated experiments have been done to quantify the tunnel coupling tunability of similar devices in QSOI\textsuperscript{\textregistered}, the result of which can be found in~\cite{farnaud_iedm_2026}.
With this in hand, our next step is to develop an automated routine to optimize the charge detection measurement sensibility.

\section{Conclusions}\label{sec:conclusions}
This work presented MOS and QD results obtained on a dedicated fabrication process flow in development for optimized 28nm FD-SOI quantum dots, establishing QSOI\textsuperscript{\textregistered} as a robust technological baseline for qubit device integration.
The implemented process improvements resulted in enhanced electrostatic control and
improved threshold voltage uniformity, both of which are critical for the reproducible formation of few-electron quantum dots.
Statistical characterization of the first-electron loading demonstrated a very small device-to-device dispersion, indicating a high degree of uniformity in quantum dot formation
and a low level of disorder across the fabricated devices.
These results confirm that our optimized process can deliver the uniform electrostatic landscape required
for scalable quantum devices.
Beyond the process development itself, the QSOI\textsuperscript{\textregistered} baseline provides a reliable platform for advanced qubit demonstrations.
In particular,
it enabled the realization of silicon spin qubits, the results of which are reported in ~\cite{farnaud_iedm_2026}.
The demonstrated process maturity and reproducibility therefore represent
an important step toward scalable 28nm FD-SOI-based quantum computing technologies.

\section*{Acknowledgment}

A special thanks to Franck Barzic and Sylvain Sausse for operating the probe stations, to Victor Doebele and Thomas Latella for the support with cryogenics,
to Matthieu Dartiailh for software support and development, and to Jean-Charles Barbé for the scientific review of this work.

\bibliographystyle{IEEEtran}
\bibliography{bibliography}

@article{gonzalez_zalba_scaling_2021,
	title = {Scaling silicon-based quantum computing using {CMOS} technology},
	volume = {4},
	issn = {2520-1131},
	url = {https://www.nature.com/articles/s41928-021-00681-y},
	doi = {10.1038/s41928-021-00681-y},
	language = {en},
	number = {12},
	urldate = {2026-05-28},
	journal = {Nature Electronics},
	author = {Gonzalez-Zalba, M. F. and De Franceschi, S. and Charbon, E. and Meunier, T. and Vinet, M. and Dzurak, A. S.},
	month = dec,
	year = {2021},
	pages = {872--884},
}

@article{maurand_cmos_2016,
	title = {A {CMOS} silicon spin qubit},
	volume = {7},
	issn = {2041-1723},
	url = {https://www.nature.com/articles/ncomms13575},
	doi = {10.1038/ncomms13575},
	language = {en},
	number = {1},
	urldate = {2026-05-28},
	journal = {Nature Communications},
	author = {Maurand, R. and Jehl, X. and Kotekar-Patil, D. and Corna, A. and Bohuslavskyi, H. and Laviéville, R. and Hutin, L. and Barraud, S. and Vinet, M. and Sanquer, M. and De Franceschi, S.},
	month = nov,
	year = {2016},
	pages = {13575},
}

@article{zwanenburg_silicon_2013,
	title = {Silicon quantum electronics},
	volume = {85},
	copyright = {http://link.aps.org/licenses/aps-default-license},
	issn = {0034-6861, 1539-0756},
	url = {https://link.aps.org/doi/10.1103/RevModPhys.85.961},
	doi = {10.1103/RevModPhys.85.961},
	language = {en},
	number = {3},
	urldate = {2026-05-28},
	journal = {Reviews of Modern Physics},
	author = {Zwanenburg, Floris A. and Dzurak, Andrew S. and Morello, Andrea and Simmons, Michelle Y. and Hollenberg, Lloyd C. L. and Klimeck, Gerhard and Rogge, Sven and Coppersmith, Susan N. and Eriksson, Mark A.},
	month = jul,
	year = {2013},
	pages = {961--1019},
}

@article{meunier_silicon_2025,
	title = {Silicon spin qubits: a viable path towards industrial manufacturing of large-scale quantum processors},
	volume = {61},
	issn = {1434-601X},
	shorttitle = {Silicon spin qubits},
	url = {https://link.springer.com/10.1140/epja/s10050-025-01514-8},
	doi = {10.1140/epja/s10050-025-01514-8},
	language = {en},
	number = {3},
	urldate = {2026-05-28},
	journal = {The European Physical Journal A},
	author = {Meunier, Tristan and Daval, Nicolas and Perruchot, François and Vinet, Maud},
	month = mar,
	year = {2025},
	pages = {58},
}

@article{dumoulin_stuyck_cmos_2026,
	title = {{CMOS} compatibility of semiconductor spin qubits},
	volume = {3},
	issn = {2948-1201},
	url = {https://www.nature.com/articles/s44287-026-00283-w},
	doi = {10.1038/s44287-026-00283-w},
	language = {en},
	number = {5},
	urldate = {2026-05-28},
	journal = {Nature Reviews Electrical Engineering},
	author = {Dumoulin Stuyck, Nard and Saraiva, Andre and Gilbert, Will and Cifuentes Pardo, Jesus and Li, Ruoyu and Escott, Christopher C. and De Greve, Kristiaan and Voinigescu, Sorin and Reilly, David J. and Dzurak, Andrew S.},
	month = apr,
	year = {2026},
	pages = {300--315},
}

@misc{bartee_spin_2024,
	title = {Spin {Qubits} with {Scalable} milli-kelvin {CMOS} {Control}},
	copyright = {Creative Commons Attribution 4.0 International},
	url = {https://arxiv.org/abs/2407.15151},
	doi = {10.48550/ARXIV.2407.15151},
	urldate = {2026-05-28},
	publisher = {arXiv},
	author = {Bartee, Samuel K. and Gilbert, Will and Zuo, Kun and Das, Kushal and Tanttu, Tuomo and Yang, Chih Hwan and Stuyck, Nard Dumoulin and Pauka, Sebastian J. and Su, Rocky Y. and Lim, Wee Han and Serrano, Santiago and Escott, Christopher C. and Hudson, Fay E. and Itoh, Kohei M. and Laucht, Arne and Dzurak, Andrew S. and Reilly, David J.},
	year = {2024},
	note = {Version Number: 1},
}

@article{neyens_probing_2024,
	title = {Probing single electrons across 300-mm spin qubit wafers},
	volume = {629},
	issn = {0028-0836, 1476-4687},
	url = {https://www.nature.com/articles/s41586-024-07275-6},
	doi = {10.1038/s41586-024-07275-6},
	language = {en},
	number = {8010},
	urldate = {2026-05-28},
	journal = {Nature},
	author = {Neyens, Samuel and Zietz, Otto K. and Watson, Thomas F. and Luthi, Florian and Nethwewala, Aditi and George, Hubert C. and Henry, Eric and Islam, Mohammad and Wagner, Andrew J. and Borjans, Felix and Connors, Elliot J. and Corrigan, J. and Curry, Matthew J. and Keith, Daniel and Kotlyar, Roza and Lampert, Lester F. and Madzik, Mateusz T. and Millard, Kent and Mohiyaddin, Fahd A. and Pellerano, Stefano and Pillarisetty, Ravi and Ramsey, Mick and Savytskyy, Rostyslav and Schaal, Simon and Zheng, Guoji and Ziegler, Joshua and Bishop, Nathaniel C. and Bojarski, Stephanie and Roberts, Jeanette and Clarke, James S.},
	month = may,
	year = {2024},
	pages = {80--85},
}

@misc{hamonic_foundry-fabricated_2025,
	title = {A foundry-fabricated spin qubit unit cell with in-situ dispersive readout},
	copyright = {Creative Commons Attribution 4.0 International},
	url = {https://arxiv.org/abs/2504.20572},
	doi = {10.48550/ARXIV.2504.20572},
	urldate = {2026-05-28},
	publisher = {arXiv},
	author = {Hamonic, Pierre and Toubeix, Mathieu and Haas, Guillermo and Nath, Jayshankar and Dartiailh, Matthieu C. and Martinez, Biel and Bertrand, Benoit and Niebojewski, Heimanu and Vinet, Maud and Bäuerle, Christopher and Balestro, Franck and Meunier, Tristan and Urdampilleta, Matias},
	year = {2025},
	note = {Version Number: 1},
}

@inproceedings{cardoso_paz_fdsoi_2024,
	address = {San Francisco, CA, USA},
	title = {{FDSOI} {Platform} for {Quantum} {Computing}},
	copyright = {https://doi.org/10.15223/policy-029},
	isbn = {979-8-3503-6542-9},
	url = {https://ieeexplore.ieee.org/document/10873521/},
	doi = {10.1109/IEDM50854.2024.10873521},
	urldate = {2026-05-28},
	booktitle = {2024 {IEEE} {International} {Electron} {Devices} {Meeting} ({IEDM})},
	publisher = {IEEE},
	author = {Cardoso Paz, B. and Elbaz, G. A. and Ouvrier-Buffet, M. and Cassé, M. and Bergamaschi, F. E. and Filippini, J.B. and Berru, J. J. Suarez and Julliard, P. L. and Diaz, B. Martinez I and Klemt, B. and El-Homsy, V. and Champain, V. and Millory, V. and Lethiecq, R. and Labracherie, V. and Roussely, G. and Bertrand, B. and Niebojewski, H. and Badets, F. and Urdampilleta, M. and De Franceschi, S. and Meunier, T. and Vinet, M.},
	month = dec,
	year = {2024},
	pages = {1--4},
}

@article{zajac_resonantly_2018,
	title = {Resonantly driven {CNOT} gate for electron spins},
	volume = {359},
	issn = {0036-8075, 1095-9203},
	url = {https://www.science.org/doi/10.1126/science.aao5965},
	doi = {10.1126/science.aao5965},
	language = {en},
	number = {6374},
	urldate = {2026-05-28},
	journal = {Science},
	author = {Zajac, D. M. and Sigillito, A. J. and Russ, M. and Borjans, F. and Taylor, J. M. and Burkard, G. and Petta, J. R.},
	month = jan,
	year = {2018},
	pages = {439--442},
}

@article{scappucci_germanium_2020,
	title = {The germanium quantum information route},
	volume = {6},
	issn = {2058-8437},
	url = {https://www.nature.com/articles/s41578-020-00262-z},
	doi = {10.1038/s41578-020-00262-z},
	language = {en},
	number = {10},
	urldate = {2026-05-28},
	journal = {Nature Reviews Materials},
	author = {Scappucci, Giordano and Kloeffel, Christoph and Zwanenburg, Floris A. and Loss, Daniel and Myronov, Maksym and Zhang, Jian-Jun and De Franceschi, Silvano and Katsaros, Georgios and Veldhorst, Menno},
	month = dec,
	year = {2020},
	pages = {926--943},
}

@article{veldhorst_two_qubit_2015,
	title = {A two-qubit logic gate in silicon},
	volume = {526},
	issn = {0028-0836, 1476-4687},
	url = {https://www.nature.com/articles/nature15263},
	doi = {10.1038/nature15263},
	language = {en},
	number = {7573},
	urldate = {2026-05-28},
	journal = {Nature},
	author = {Veldhorst, M. and Yang, C. H. and Hwang, J. C. C. and Huang, W. and Dehollain, J. P. and Muhonen, J. T. and Simmons, S. and Laucht, A. and Hudson, F. E. and Itoh, K. M. and Morello, A. and Dzurak, A. S.},
	month = oct,
	year = {2015},
	pages = {410--414},
}

@article{elbaz_integration_2025,
	title = {Integration of {W} vias for individual coupling control in 28 nm {FD}-{SOI} qubit arrays},
	volume = {229},
	issn = {00381101},
	url = {https://linkinghub.elsevier.com/retrieve/pii/S0038110125001509},
	doi = {10.1016/j.sse.2025.109205},
	language = {en},
	urldate = {2026-06-11},
	journal = {Solid-State Electronics},
	author = {Elbaz, G.A. and Pelloux-Prayer, J. and Gruel, K. and Torresani, P. and Lethiecq, R. and Julliard, P.L. and Suarez-Segovia, C. and Arnaud, F. and Nowak, E. and Meunier, T. and Paz, B.C.},
	month = nov,
	year = {2025},
	pages = {109205},
}

@article{zwolak_data_2024,
	title = {Data needs and challenges for quantum dot devices automation},
	volume = {10},
	issn = {2056-6387},
	url = {https://www.nature.com/articles/s41534-024-00878-x},
	doi = {10.1038/s41534-024-00878-x},
	language = {en},
	number = {1},
	urldate = {2026-07-21},
	journal = {npj Quantum Information},
	author = {Zwolak, Justyna P. and Taylor, Jacob M. and Andrews, Reed W. and Benson, Jared and Bryant, Garnett W. and Buterakos, Donovan and Chatterjee, Anasua and Das Sarma, Sankar and Eriksson, Mark A. and Greplová, Eliška and Gullans, Michael J. and Hader, Fabian and Kovach, Tyler J. and Mundada, Pranav S. and Ramsey, Mick and Rasmussen, Torbjørn and Severin, Brandon and Sigillito, Anthony and Undseth, Brennan and Weber, Brian},
	month = oct,
	year = {2024},
	pages = {105},
}

@article{farnaud_iedm_2026,
	author = {
	F. Arnaud and B. Cardoso Paz and E. Nowak and C.A. Suarez-Segovia and R. Duru and J.G.
Simiz and C. Renard and S. Nneme and B.Pernet and A. Pisanu and C. Jenny and L. Parmigiani and
S. Lagrasta and O. Gourhant and C. Pribat and P. Garnier and F. Foussadier and A. Durel and M.
Gregoire and M.G. Gusmao-Cacho and N. Potrzebowska and A. Moroni and M. Repossi and
L. Gerosa and J.D. Chapon and E. Rouchouze and L. Favennec and R. Bouyssou and M.
Gala-el-dine and P. Galy and F. Sonnerat and B. Dormieu and D. Benoit and A. Sangnier and
C. Duluard and P. Torresani and G.A. Elbaz and A. Comiti and F. Barzic and S. Sausse and
M. Brzezinska and J. Guigon and N. Kubler and B. Vermersch and J. Pelloux-Prayer and E.
Prin and O. Daoudi and L. Bresque and P. Lheritier and E. Plouet and A.A. Aravindnath and T.
Aladjidi and J.C. Barbé and G. Britton and C. Condemine and A.C. Fernandez-Rodas and
M. Melo-de-Lima and F. Sabatini and A. Lepage and A. Ronco and A. Poulet and M. A.
Usuga and S. Martin and M. Dartiailh and T. Latella and V. Doebele and C. Eymard and C.
Sejor and P. Pedram and C. Lelez and P.L. Julliard and K. Gruel and M. Darmyn and C. Le Lez and
R. Lethiecq and D. Huynh and N. Daval and I. Sidibe and M. Alepidis and T. Bedecarrats and
J. Lugo-Alvarez and M. Cassé and A. Bajolet and V. Sabia-Pereira-Carpes and F.E.
Bergamaschi and M. Vinet and T. Meunier
},
	title = {28 nm FD-SOI based qubit/logic
300 mm platform for quantum processor unit chip compatible with volume
manufacturing},
	journal = {IEDM},
	year = {to be presented at IEDM 2026, (accepted)},
}

@inproceedings{mazurier_variability_2014,
	title = {Variability of planar {Ultra}-{Thin} {Body} and {Buried} oxide ({UTBB}) {FDSOI} {MOSFETs}},
	doi = {10.1109/ICICDT.2014.6838617},
	booktitle = {2014 {IEEE} {International} {Conference} on {IC} {Design} {Technology}},
	author = {Mazurier, J. and Weber, O. and Andrieu, F. and Royer, C. L. and Faynot, O. and Vinet, M.},
	month = may,
	year = {2014},
	pages = {1--4},
}

\end{document}